\documentclass[10pt,letterpaper,conference]{IEEEtran}
\IEEEoverridecommandlockouts

\usepackage{amsmath}
\usepackage{graphicx}
\usepackage{booktabs}
\usepackage{url}
\usepackage{tabularx}
\usepackage{lipsum} 
\usepackage{balance}
\usepackage{enumitem}
\usepackage{caption}
\usepackage{subcaption}
\usepackage{multirow}
\usepackage{array}
\usepackage{amsmath}
\usepackage{diagbox}
\usepackage{slashbox}
\usepackage{amsfonts} 
\usepackage{xcolor} 
\usepackage{graphicx} 
\usepackage{comment} 
\usepackage{amssymb} 
\usepackage{subcaption} 

\usepackage[numbers,sort&compress]{natbib} 

\usepackage[left=0.65in,right=0.65in,top=0.75in,bottom=1.1in]{geometry}
\begin{document}

\title{Indoor/Outdoor Spectrum Sharing Enabled by GNSS-based Classifiers}

\author{
\IEEEauthorblockN{
Hossein Nasiri,
Muhammad Iqbal Rochman,
and Monisha Ghosh}
\IEEEauthorblockA{
Department of Electrical Engineering, University of Notre Dame \\ 
Email: \{hnasiri2,mrochman,mghosh3\}@nd.edu}
}

\maketitle

\begin{abstract}

The desirability of the mid-band frequency range (1 - 10 GHz) for federal and commercial applications, combined with the growing applications for commercial indoor use-cases, such as factory automation, opens up a new approach to spectrum sharing: the same frequency bands used outdoors by federal incumbents can be reused by commercial indoor users. A recent example of such sharing, between commercial systems, is the 6 GHz band (5.925 - 7.125 GHz) where unlicensed, low-power-indoor (LPI) users share the band with outdoor incumbents, primarily fixed microwave links. However, to date, there exist no reliable, automatic means of determining whether a device is indoors or outdoors, necessitating the use of other mechanisms such as mandating indoor access points (APs) to have integrated antennas and not be battery powered, and reducing transmit power of client devices which may be outdoors.
An accurate indoor/outdoor (I/O) classification addresses these challenges, enabling automatic transmit power adjustments without interfering with incumbents.
To this end, we leverage the Global Navigation Satellite System (GNSS) signals for I/O classification. GNSS signals, designed inherently for outdoor reception and
highly susceptible to indoor attenuation and blocking, provide
a robust and distinguishing feature for environmental sensing. 
We develop various methodologies, including threshold-based techniques and machine learning approaches and evaluate them using an expanded dataset gathered from diverse geographical locations. Our results demonstrate that GNSS-based methods alone can achieve greater accuracy than approaches relying solely on wireless (Wi-Fi) data, particularly in unfamiliar locations. Furthermore, the integration of GNSS data with Wi-Fi information leads to improved classification accuracy, showcasing the significant benefits of multi-modal data fusion.

\end{abstract}

\section{Introduction}

The ongoing expansion of wireless communication continues to strain available spectrum, particularly in the mid-band range (1–10 GHz). To address this, regulators have increasingly adopted spectrum sharing models that promote coexistence between licensed and unlicensed devices. A notable example is the allocation of the 6 GHz band (5.925–7.125 GHz) for unlicensed use, a band that includes many incumbent services such as fixed microwave links~\cite{2025_6GHz_CommMag}. 
Given the potential for harmful interference with outdoor incumbents, the Federal Communications Commission (FCC) permits unlicensed Low Power Indoor (LPI) access points (APs) to operate without Automated Frequency Coordination (AFC) under strict indoor-only constraints. Indoor transmissions experience significant attenuation and scattering due to building materials, which substantially reduces the likelihood of interference with outdoor incumbent systems. By limiting operation to indoor environments, regulators can safely expand shared access, if devices can be reliably classified as operating indoors.
Clients in the 6 GHz band typically connect to LPI APs, but the FCC is also considering enabling a new mode of operation called Client-to-Client (C2C) communication, where client devices can directly exchange data after receiving an enabling signal from an LPI AP. Unlike LPI APs, however, client devices are not inherently restricted to indoor use, making it riskier to allow them to operate at equivalent transmit power levels. In this context, robust and accurate indoor/outdoor (I/O) classification becomes essential, not only to enforce indoor-only constraints but also to manage client transmit power in a way that protects incumbents while maximizing unlicensed utility. Robust I/O classification methodologies can also enable sharing with co-channel or adjacent channel federal incumbents, for example, in the 3.1 - 3.45 GHz band, or the upper C-band, 3.98 - 4.2 GHz~\cite{FCC_UpperC-band_NOI} where adjacent-channel interference in radar altimeters is a concern.

In previous work, I/O classification has been explored using cellular and Wi-Fi signals. However, these methods are limited, primarily due to their reliance on cellular and Wi-Fi infrastructure, which may not be consistently available in all locations, and limited generalization of performance across diverse geographical areas. This paper addresses these challenges by incorporating measurements from the Global Navigation Satellite System (GNSS) directly into the I/O classification process: note that we do not simply process the localization information from the device, which can be extremely inaccurate, especially indoors. GNSS offers significant advantages, including global availability (signals can be received in most locations worldwide) and inexpensive receivers that are available even in small form-factor devices such as watches. Furthermore, the known and relatively small number of GNSS satellites simplifies understanding their behavior.

The main contributions of this paper are fourfold. First, we extended the SigCap Android application~\cite{sigcap} to enable GNSS-enhanced data collection with detailed satellite-specific 
measurements. Second, we built an expanded dataset covering multiple cities in the USA as well as several international locations, thereby addressing the geographic limitations of prior studies. Third, we carried out a comparative analysis of different approaches, including threshold-based methods and machine learning (ML) models, for indoor/outdoor classification using this dataset. Finally, we show that incorporating GNSS data leads to improved accuracy and generalization, particularly in challenging scenarios.

The rest of this paper is organized as follows: Section~2 reviews related work, 
Section~3 describes the data collection and preprocessing methodology, Section~4 presents the 
feature analysis and classification methods, Section~5 reports the experimental 
results, Section~6 discusses practical considerations and Section~7 concludes the 
paper with implications of our findings and directions for future research.

\section{Related Work}

Previous research has explored I/O classification using wireless signals. Our prior work \cite{ACMPaper} leveraged cellular and Wi-Fi signals to classify devices into three categories: indoor-interior, indoor-near-window, and outdoor. We showed that both classical and deep learning models can achieve relatively accurate predictions. However, this work has some limitations as discussed, primarily related to the dynamic nature of cellular and Wi-Fi deployments. 

Several other studies have investigated I/O classification using various techniques. Lee et al.~\cite{lee-gps-temp} combined GNSS and temperature data for personal exposure assessment. Although this work achieved a balanced accuracy of 90\%, it presented several key limitations: the data lacked geographical diversity, being collected solely in Seoul, South Korea; the collection procedure employed a highly regimented routine (e.g., 15 minutes walking indoors followed by 15 minutes outdoors), which fails to simulate the varied motion patterns of real-world mobile devices; and its reliance on a temperature sensor, a component not commonly found in standard wireless devices like mobile phones, significantly reduces its practical applicability for spectrum sharing. While Bui et al.~\cite{bui-gps-ml} proposed a GNSS-based I/O detection scheme using ML, their work suffered from slow data updates and a lack of geographical diversity, as data was collected in only one specific location and its surrounding area, thereby limiting the evaluation of the proposed methods' generalization capability. Zhang et al.~\cite{zhang-ensemble} utilized an ensemble learning scheme and Long-Term Evolution (LTE) signals, collected via a specific mobile application, for I/O classification. Similarly, this work suffered from two key limitations such as not providing a comprehensive evaluation of its dataset across different geographical locations, and being restricted to devices equipped with cellular networks. Ramamurthy et al.~\cite{ramamurthy2022ml} incorporated both Wi-Fi and cellular data to train various ML methods, which were then evaluated across multiple locations. However, the models failed to maintain high accuracy when tested on previously unseen locations.

Our paper differentiates itself by incorporating detailed information for each satellite in the four major GNSS constellations—GPS (United States), GLONASS (Russia), Galileo (European Union), and BeiDou (China)—and by using a geographically diverse dataset collected in dynamic, real-world scenarios.

\section{Dataset Overview}
A central contribution of this study is the development of a geographically diverse dataset for I/O classification using GNSS signals. This section details the dataset from collection stages to preprocessing, and preparation for subsequent modeling.

\subsection{GNSS Data Collection}
\begin{figure}[t]
    \centering
    \includegraphics[width=0.48\textwidth]{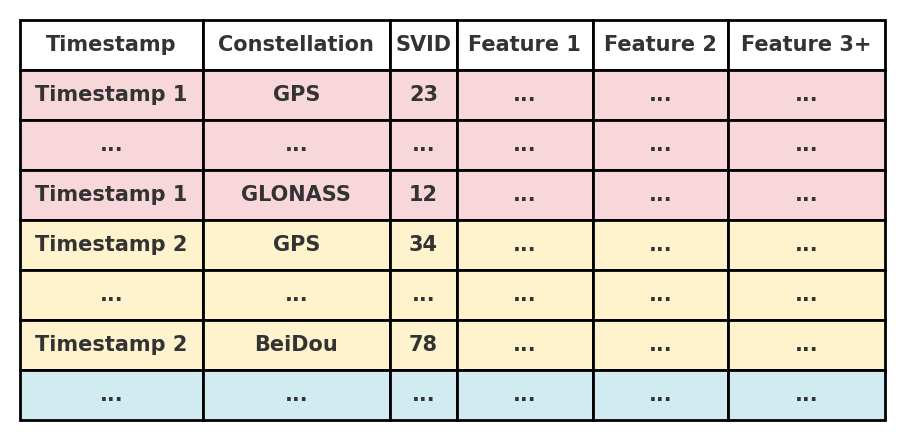}
    \caption{Symbolic illustration of the dataset structure. Rows with the same timestamp are grouped with a shared background color.}
    \label{fig:data_struct}
\end{figure}

We collected the necessary data using a modified version of the SigCap Android application \cite{sigcap}, which now supports GNSS data recording in addition to its existing Wi-Fi and cellular measurement capabilities. The dataset was gathered under dynamic conditions, with outdoor measurements obtained while walking or driving, and indoor measurements collected while standing or walking slowly. SigCap records data every 5 seconds. At each timestamp, the device may observe from 0 up to more than 50 GNSS satellites. For each observed satellite, SigCap records detailed information including basic metadata (timestamp, Space Vehicle Identification (SVID), constellation type), angular parameters (azimuth, elevation), and signal characteristics (carrier frequency, carrier-to-noise ratio (CNR)). A summary of these GNSS features is provided in Table~\ref{tab:gps_data_features}. To further clarify the structure of the GNSS dataset (exported as CSV files), we also include a symbolic illustration of a single GNSS CSV file in Figure~\ref{fig:data_struct}. A single start–stop session of SigCap produces one CSV file. At each location, we divided measurements into shorter sessions (more than 95\%  of files are under 60 minutes) to reduce the impact of time-varying factors (e.g., weather) that could affect the recorded features. Consequently, multiple CSV files were generated, each with appropriate labeling and description. During each session, the environment (indoor or outdoor) was kept constant. Thus, all rows in a given CSV file correspond to a single label.  

 Beyond GNSS, the application additionally collects cellular and Wi-Fi measurements such as Reference Signal Received Power (RSRP), Reference Signal Received Quality (RSRQ), Received Signal Strength Indicator (RSSI), and frequency information.  

\begin{table}[h!]
\centering
\caption{GNSS data recorded by SigCap for each satellite.}
\label{tab:gps_data_features}
\resizebox{\linewidth}{!}{
\begin{tabular}{|l|l|} 
\hline
\textbf{Category} & \textbf{Feature} \\ \hline
\multirow{3}{*}{Basic Information} & Timestamp \\
                                 & Satellite ID (SVID) \\
                                 & Constellation type \\ \hline
\multirow{2}{*}{Angular Information} & Azimuth \\
                                     & Elevation \\ \hline
\multirow{2}{*}{Signal Information}  & Carrier frequency \\
                                 & CNR \\ \hline
\multirow{3}{*}{Additional Data}   & Almanac and ephemeris data \\
                                 & Usage in fix \\
                         \hline
\end{tabular} 
}
\end{table}

To mitigate any bias towards specific phone brands, the data collection used a range of different mobile brands and models: Google Pixel 5, Google Pixel 6, Google Pixel 8, Samsung S21, Samsung S22, and Samsung S24.

\subsection{Dataset Summary and Geographic Diversity}

\begin{figure}[t]
    \centering
    \includegraphics[width=0.96\columnwidth]{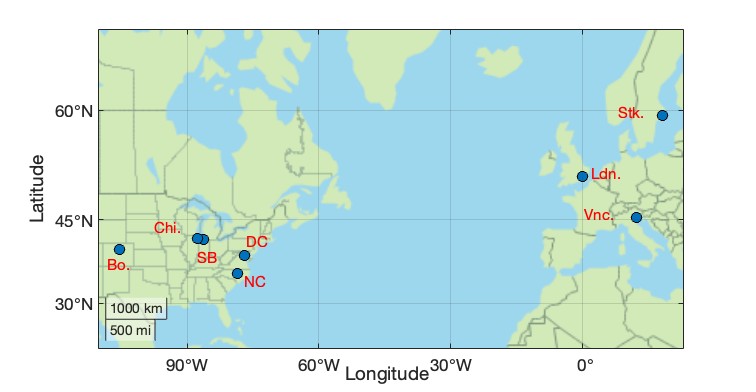} 
    \caption{Geographic locations used in data collection (see Table~\ref{tab:indoor_outdoor_samples} for abbreviations).}
    \label{fig:dataset_map}
\end{figure}

The dataset collection, illustrated in Figure~\ref{fig:dataset_map}, took place across multiple locations in the USA as well as selected countries in Europe. The map highlights the geographical diversity of our dataset, addressing one of the major limitations of prior I/O classification studies, namely the lack of environmental variety. This diversity enables us to evaluate proposed methods across a wide range of environments and conditions.
Table \ref{tab:indoor_outdoor_samples} also summarizes the dataset sizes for each location, categorized by their corresponding environment (indoor or outdoor). For experimental purposes, we further divide the datasets into two groups: Group A and Group B. Group A datasets are used for both training and testing. In contrast, Group B datasets are reserved exclusively for testing, allowing us to assess the generalizability of our models to unseen environments.

\begin{table}[h!]
\caption{Number of indoor and outdoor samples per location.}
\centering
\resizebox{\columnwidth}{!}{%
\begin{tabular}{lrrp{0.8cm}} 
\hline
\textbf{Location} & \textbf{Indoor} & \textbf{Outdoor} & \textbf{Grp} \\  
\hline
South Bend (SB)        & 107291 & 110432 & A \\
Chicago (Chi.)         & 71825  & 109312 & A \\
Washington, D.C. (DC)  & 19698  & 15844  & A \\
Boulder (Bo.)          & 6850   & 11538  & A \\
North Carolina (NC)    & 21760  & 0      & A \\
Stockholm (Stk.)       & 695    & 870    & A \\
London (Ldn.)          & 16673  & 101075 & B \\
Venice (Vnc.)          & 20855  & 10866  & B \\
\hline
\end{tabular}}
\label{tab:indoor_outdoor_samples}
\end{table}

\subsection{Data Processing and Preparation}
The raw dataset described earlier cannot be directly used for training and evaluation. It requires thorough cleaning, processing, and transformation to ensure that imperfections common in real-world datasets do not hinder model training or prediction accuracy. The refinement procedure includes:

\subsubsection*{1) Cleaning the Dataset}
Datasets might contain invalid or missing values due to software or hardware errors. In our case, one common anomaly was the observation of satellites with a CNR equal to 0 dB/Hz, which is far below the typical minimum detection threshold of approximately 10 dB/Hz for GNSS receivers \cite{paziewski2019signal}. These anomalous entries also contained missing frequency values. We observed that this phenomenon occurred primarily with older devices (e.g., Google Pixel 5) and is likely residual information carried over from previous timestamps. Filtering such values is critical, since CNR (as shown in later sections) is one of the most important features for I/O classification, carrying significant information about environmental isolation. 

In addition, we discarded the first 20 seconds of data from each measurement file. This was necessary to allow GNSS receivers sufficient time to initialize and acquire stable positional information after startup.

\subsubsection*{2) Grouping GNSS Data}
For specific classification methods, the GNSS data underwent grouping based on key satellite characteristics. This included grouping by constellation type (e.g., GPS, GLONASS, Galileo, and BeiDou), SVID, and frequency of operation.

\subsubsection*{3) Normalization}
Given the diverse range and varying scales of features within our dataset, it was essential to normalize the data. Prior to normalization, categorical attributes such as constellation type were converted into numeric representations to allow uniform processing. We then employed Min-Max normalization to scale all numerical values in each column to the range $[0,1]$. This step ensures that features with inherently larger numerical ranges do not disproportionately influence the training of ML models, thereby promoting balanced learning across all features.

\section{Feature Analysis and Classifiers}

\subsection{Feature Analysis}
\label{FAn}
In this section, we present a few representative analyses that best illustrate how different GNSS features behave under indoor and outdoor conditions, providing insight into their discriminatory power.

\subsubsection{Receiver Operating Characteristic (ROC) Curves}
\label{sec:roc}
\begin{figure*}[h!]
    \centering
    \begin{subfigure}[b]{0.32\textwidth}
        \centering
        \includegraphics[width=\linewidth]{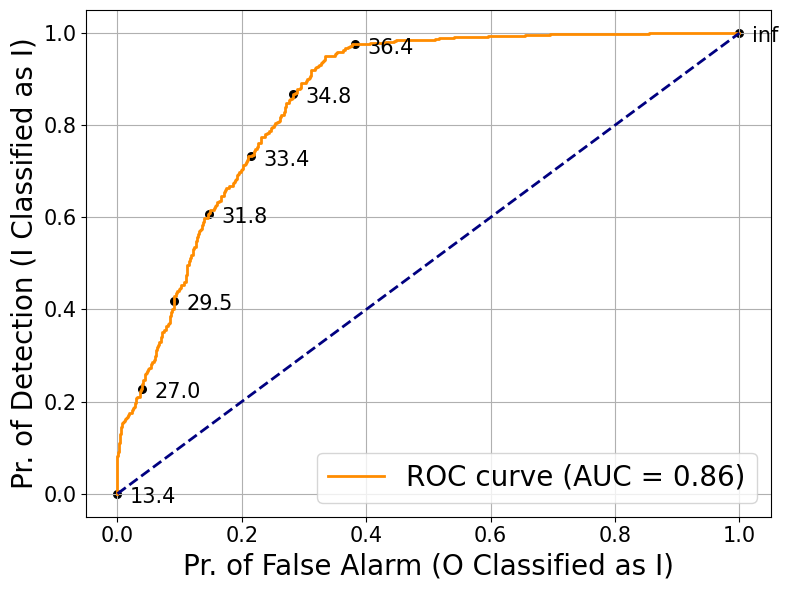}
        \caption{ROC curve of CNR values for BeiDou, SVID: 26, Carrier Frequency: 1575.4 MHz.}
        \label{fig:roc_individual_satellite_example}
    \end{subfigure}
    \hfill
    \begin{subfigure}[b]{0.32\textwidth}
        \centering
        \includegraphics[width=\linewidth]{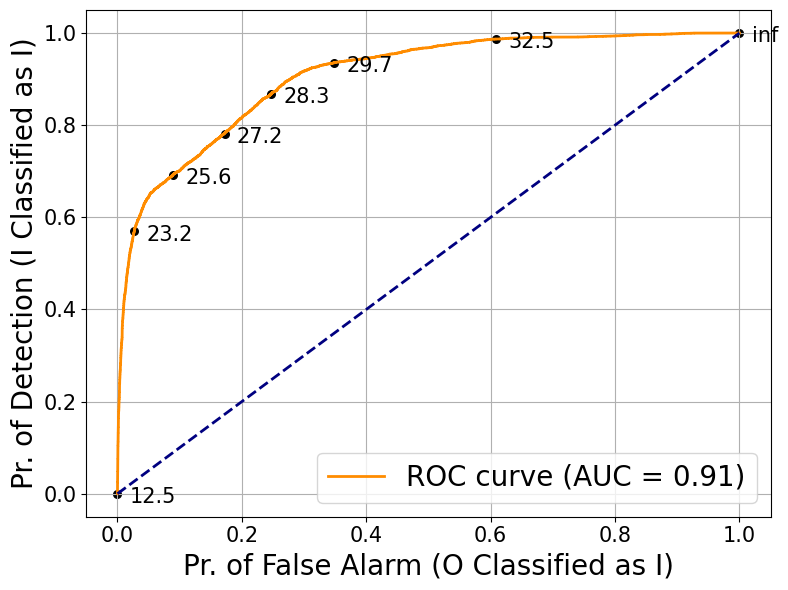}
        \caption{ROC curve of the mean CNR value per timestamp.}
        \label{fig:roc_mean_cnr}
    \end{subfigure}
    \hfill
    \begin{subfigure}[b]{0.32\textwidth}
        \centering
        \includegraphics[width=\linewidth]{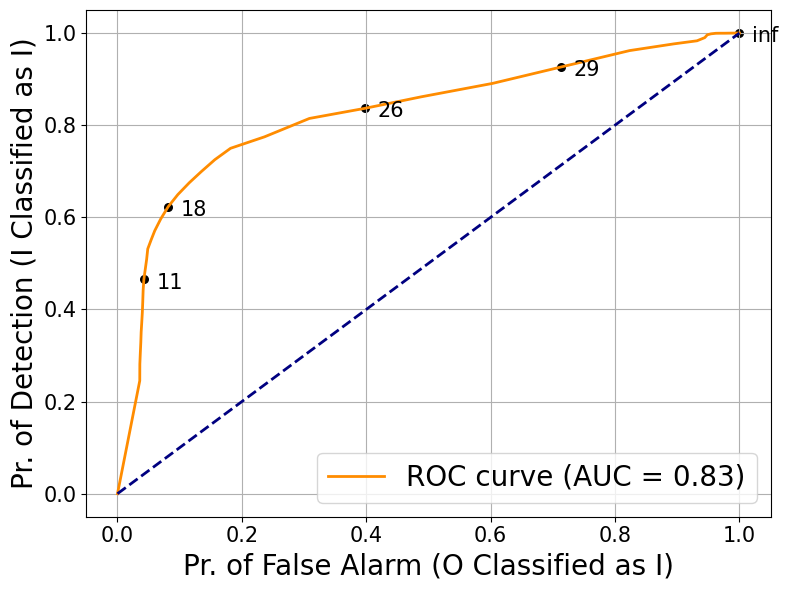}
        \caption{ROC curve of the number of satellites per timestamp.}
        \label{fig:num_sat_cnr}
    \end{subfigure}

    \caption{ROC curves for (a) an individual satellite's CNR, (b) mean CNR (c) number of observed satellites.}
    \label{fig:roc}
\end{figure*}

An ROC curve is a graphical plot that illustrates the diagnostic ability of a binary classifier system as its discrimination threshold is varied.
Figure \ref{fig:roc} illustrates the ROC plot of different features. Specifically, Figure \ref{fig:roc_individual_satellite_example} displays the ROC curve for CNR values from a specific satellite (a BeiDou satellite with SVID 26 operating at 1575.4 MHz). Here, "detection" is defined as correctly classifying an indoor instance as indoor, and "false alarm" as misclassifying an outdoor instance as indoor. The figure demonstrates that by setting a CNR threshold of approximately 33.4 dB/Hz, around 75\% of indoor data can be correctly detected, while only 20\% of outdoor data is misclassified.

Furthermore, Figure \ref{fig:roc_mean_cnr} presents the ROC curve for the linear average of CNR values calculated across all satellites observed at each timestamp. This aggregation enhances performance; if a threshold of 27.2 dB/Hz is applied to the mean CNR, approximately 80\% of indoor data is correctly detected, with only 15\% of outdoor data being misclassified. This highlights that even simple averaging of CNR values, without separating individual satellites, can result in substantially improved classification accuracy.

Just as CNR tends to be higher outdoors, the total number of observed satellites at any time also tends to increase in outdoor settings. Figure \ref{fig:num_sat_cnr} clearly shows this. For instance, we saw more than 18 satellites in about 40\% of indoor observations, but this happened in nearly 90\% of outdoor observations.

\subsubsection{CNR vs. Elevation Angle Scatter Plots}
\begin{figure}[h!]
    \centering
    \begin{subfigure}[b]{\columnwidth} 
        \centering
        \includegraphics[width=0.7\columnwidth]{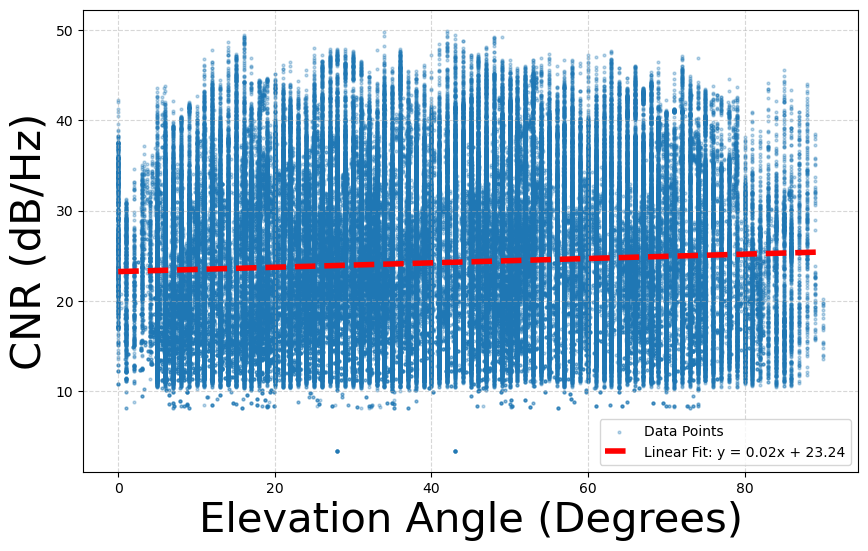} 
        \caption{Indoor environments}
        \label{fig:cnr_elevation_indoor}
    \end{subfigure}


    \begin{subfigure}[b]{\columnwidth} 
        \centering
        \includegraphics[width=0.7\columnwidth]{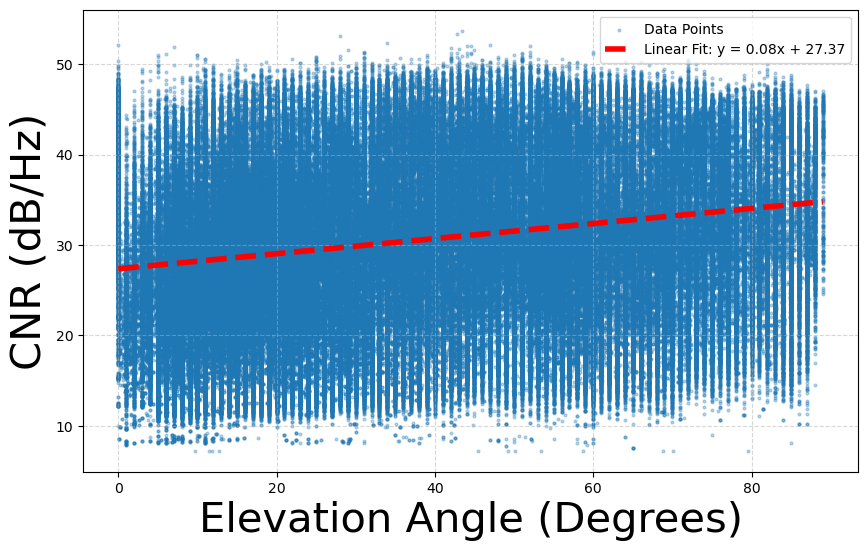} 
        \caption{Outdoor environments}
        \label{fig:cnr_elevation_outdoor}
    \end{subfigure}
    \caption{Scatter plots of CNR vs. Elevation Angle.}
    \label{fig:cnr_elevation_combined}
\end{figure}

Figure \ref{fig:cnr_elevation_combined} displays scatter plots of CNR versus elevation angle for both indoor and outdoor environments. In outdoor settings, there is generally a stronger correlation between a satellite's elevation angle and its CNR. This is because signals arriving directly and vertically from higher in the sky are typically the strongest and least obstructed.
This behavior changes notably in indoor environments, where GNSS signals are heavily attenuated and scattered by building materials. As illustrated in Figures \ref{fig:cnr_elevation_combined}, the distinction between indoor and outdoor CNR becomes more pronounced at higher elevation angles. This observable divergence at higher elevation angles inspires us to consider the elevation angle as a valuable feature for I/O classification.

\subsection{Classification Methods}
Two different types of methods were utilized for the classification task as described below:

\subsubsection*{a) Threshold-Based (Th-Based) Method}
This approach classifies indoor or outdoor environments by comparing the satellite's CNR level against a predefined \textbf{threshold}. Each unique satellite-identified by its constellation type, SVID, and carrier frequency-has its own corresponding threshold.

We first gather all available satellite information from the training set. For each satellite, we construct a table to evaluate performance metrics across a range of arbitrary thresholds. This table provides:

\noindent $\bullet$
\textbf{Probability of Detection (PD):} Defined as the percentage of \emph{indoor data samples correctly classified as indoor}.

\noindent $\bullet$
\textbf{Probability of False Alarm (PF):} Defined as the percentage of \emph{outdoor data samples misclassified as indoor}.

This process generates a table of PD and PF values for various thresholds for each satellite. Since there is a natural trade-off between PD and PF, we select the optimal threshold that maximizes the overall prediction accuracy calculated using the formula:
\begin{align*}
\text{Total Accuracy} = {} & \frac{(\text{PD} \times N_{indoor})}{N_{total}} \\
& {} + \frac{((1 - \text{PF}) \times N_{outdoor})}{N_{total}}
\end{align*}
This process yields a specific threshold for each satellite when applied to the training set.
We then evaluate new data by comparing the measured CNR of each detected satellite with its corresponding optimal threshold stored in a lookup table: if the CNR is greater than the threshold, the prediction for that satellite is outdoor; otherwise, it is indoor. Finally, the majority voting rule determines the final class label 
for each timestamp by aggregating the predictions of all 
per-satellite classifiers. However, if the number of observed 
satellites at a given timestamp is less than or equal to 10, the 
environment is automatically classified as \emph{indoor}. This is 
motivated by the ROC analysis in Fig.~\ref{fig:num_sat_cnr}, which shows that 
timestamps with 10 or fewer visible satellites are very unlikely 
to correspond to outdoor environments. 



\subsubsection*{b) ML-Based Method}
\label{sec:ml-method}
For the ML models, we treated each satellite observation as an individual entry for training. All features listed in Table~\ref{tab:gps_data_features}, except for Timestamp, Almanac, and Ephemeris data, were used as input features. In addition, for each satellite we incorporated two important cross-information features: \textbf{average CNR} and the \textbf{total number of observed satellites} at the corresponding timestamp. As shown in the section \ref{FAn}, these two features provide crucial context about the overall satellite signal environment and, by definition, remain fixed across all satellite entries within the same timestamp. Similar to the Th-based model, at the prediction stage, the final label for a timestamp is determined by majority voting.

The following ML algorithms were employed in our ML-based classification approach:

\noindent $\bullet$
\textbf{Support Vector Machine (SVM):} This model works by finding the optimal hyperplane that best separates data points belonging to different classes in a high-dimensional space.

\noindent $\bullet$
\textbf{Decision Tree (DT):} A decision tree makes predictions by learning simple decision rules inferred from the data features, forming a tree-like structure of choices.

\noindent $\bullet$
\textbf{Random Forest (RF):} An ensemble learning method, Random Forest improves accuracy and reduces overfitting by constructing multiple decision trees during training and outputting the mode of the classes (classification) or the mean prediction of the individual trees.

\section{Results}
To evaluate our models, we considered two scenarios: \noindent\textbf{Scenario 1 – Mixed Environment Testing:} Group A's data (see Table \ref{tab:indoor_outdoor_samples}) are partitioned by reserving 20\% for testing and using the remaining 80\% for training. Consequently, our models were tested on data encompassing both \emph{known environments} (from Group A's training portion) and \emph{previously unseen environments} (from Group B's testing portion).\noindent 
\textbf{ Scenario 2 – Entirely New Environment Testing:} Group A's data are exclusively used for training, while all of Group B's data served as the dedicated testing set. This setup evaluates the ability of the models to generalize to \emph{completely new and previously unseen environments}.

\subsection{Classification Performance}

This section evaluates the performance of the implemented classification methods. Table \ref{tab:performance_5s_TH} reports the accuracy of all models, broken down by environment  type and evaluated under both testing scenarios.

\begin{table}[h!]
    \centering
    \caption{Accuracy (\%) of methods for a 5-second window.}
    \label{tab:performance_5s_TH}
    \begin{tabularx}{\columnwidth}{|p{2cm}|X|X|X|X|}
        \hline
        \textbf{Method} & \multicolumn{2}{c|}{\textbf{Scenario 1}} & \multicolumn{2}{c|}{\textbf{Scenario 2}} \\
        \cline{2-5} 
        & \textbf{I} & \textbf{O} & \textbf{I} & \textbf{O} \\ \hline
        Th-B & 72 & 79 & 67 & 86 \\ \hline
        SVM & 60 & 100 & 69 & 100  \\ \hline
        RF & 78 & 95 & 78 & 100 \\ \hline
        DT & 72 & 95 & 70 & 100 \\ \hline

    \end{tabularx}
\end{table}

A noticeable trend in the results is that outdoor accuracy 
was significantly higher than indoor accuracy. This disparity 
can be attributed to the fact that more than 50\% of our indoor 
dataset was collected near windows, where signal characteristics 
often resemble those of outdoor environments. As discussed in 
section~\ref{containment}, we explicitly analyze the impact of isolation 
independently of the I/O state. When restricting the evaluation 
to environments explicitly labeled as \emph{indoor-interior} 
(i.e., locations not near windows), the average indoor accuracy 
improved by approximately 10\% across all 
methods, with RF achieving 87\% accuracy for indoor classification. 

\subsection{Further Improvement by Temporal Aggregation (Extended Window Technique)}
\label{sec:temporal-aggregation}

The SigCap application records data every 5 seconds, which means that 
I/O state predictions can, in principle, be updated at the same rate. 
In practice, however, a device’s physical environment rarely changes 
within such short intervals. Moreover, overly frequent updates may 
produce unstable results due to minor signal fluctuations. To improve 
reliability and mitigate the impact of brief errors or noise, we apply 
\emph{temporal aggregation}, also referred to as the extended window 
technique. Instead of relying on a single 5-second prediction, 
predictions from multiple consecutive timestamps are combined into 
a larger time window (e.g., 30 or 60 seconds), providing a smoother 
and more robust estimate of the I/O state.

Tables~\ref{tab:performance_30s_TH} and~\ref{tab:performance_60s_TH} report the performance of the different classifiers 
after applying the windowing technique with lengths of 30s and 60s, respectively. 
All other model configurations remained unchanged, since the windowing method 
operates only on the prediction outputs. The results indicate a relative improvement 
when applying a 30-second window compared to the base predictor. However, extending the 
interval further from 30s to 60s yields smaller additional gains.

\begin{table}[h!]
    \centering
    \caption{Accuracy (\%) of methods for a 30-second window.}
    \label{tab:performance_30s_TH}
    \begin{tabularx}{\columnwidth}{|p{2cm}|X|X|X|X|}
        \hline
        \textbf{Method} & \multicolumn{2}{c|}{\textbf{Scenario 1}} & \multicolumn{2}{c|}{\textbf{Scenario 2}} \\
        \cline{2-5} 
        & \textbf{I} & \textbf{O} & \textbf{I} & \textbf{O} \\ \hline
        Th-B & 83 & 85 & 71 & 88 \\ \hline
        SVM & 69 & 100 & 70 & 100 \\ \hline
        RF & 79 & 99 & 80 & 100 \\ \hline
        DT & 75 & 99 & 74 & 100 \\ \hline
    \end{tabularx}
\end{table}

\begin{table}[h!]
    \centering
    \caption{Accuracy (\%) of methods for a 60-second window.}
    \label{tab:performance_60s_TH}
    \begin{tabularx}{\columnwidth}{|p{2cm}|X|X|X|X|}
        \hline
        \textbf{Method} & \multicolumn{2}{c|}{\textbf{Scenario 1}} & \multicolumn{2}{c|}{\textbf{Scenario 2}} \\
        \cline{2-5} 
        & \textbf{I} & \textbf{O} & \textbf{I} & \textbf{O} \\ \hline
        Th-B & 84 & 84 & 73 & 88 \\ \hline
        SVM & 70 & 100 & 70 & 100 \\ \hline
        RF & 80 & 98 & 80 & 100 \\ \hline
        DT & 76 & 99 & 75& 100 \\ \hline
    \end{tabularx}
\end{table}

\subsection{Comparison of GNSS and Wireless Data Classification} After evaluating the performance of methods using GNSS data, it is natural to ask how models trained on GNSS features compare to those trained on Wi-Fi  features, or whether combining both modalities can yield stronger and more 
accurate predictors. To investigate this, we conducted a performance  evaluation of our ML models using three settings: 
(1) Wi-Fi features only, 
(2) GNSS features only, and 
(3) GNSS and Wi-Fi features combined. 

The Th-based method was not considered in this section, as it is not applicable to Wi-Fi features: the large number of access points makes it impractical to extract consistent characteristics for each observed AP. For Wi-Fi, we extracted six key features aggregated within each timestamp: number of 2.4~GHz APs, number of 5~GHz APs, mean RSSI of Wi-Fi~2.4~GHz, 
mean RSSI of Wi-Fi~5~GHz, maximum RSSI of Wi-Fi~2.4~GHz, and maximum RSSI of 
Wi-Fi~5~GHz. Due to space constraints, we refer the reader to our earlier work \cite{ACMPaper} for the full details of the Wi-Fi preprocessing pipeline. 

It is also worth noting that while direct per-satellite timestamp aggregation (as in the threshold method) is not applicable to Wi-Fi features, the \textit{temporal aggregation} (extended window technique) can still be applied.

Tables~\ref{tab:wifi_results_5s}, \ref{tab:wifi_results_30s}, and \ref{tab:wifi_results_60s} 
present the comparative results for the base case (5s) as well as the 30-second and 60-second 
window settings, respectively. These results pertain to \textbf{Scenario 1}, where a mixed dataset of Group A and Group B was used for testing.

\begin{table}[h!]
    \centering
    \caption{Accuracy (\%) Comparison of Wi-Fi and GNSS features- 5-second window.}
    \label{tab:wifi_results_5s}
    \begin{tabularx}{\columnwidth}{|p{2cm}|X|X|X|X|X|X|}
        \hline
        \textbf{Method} & \multicolumn{2}{c|}{\textbf{only GNSS}} & \multicolumn{2}{c|}{\textbf{only Wi-Fi}} & \multicolumn{2}{c|}{\textbf{GNSS + Wi-Fi}} \\
        \cline{2-7}
        & \textbf{I} & \textbf{O} & \textbf{I} & \textbf{O} & \textbf{I} & \textbf{O} \\ \hline
        SVM & 60 & 100 & 54 & 84 & 69 & 99 \\ \hline
        RF & 78 & 95 & 72 & 95 & 80 & 98 \\ \hline
        DT & 72 & 95 & 70 & 95 & 80 & 98 \\ \hline
    \end{tabularx}
\end{table}

\begin{table}[h!]
    \centering
    \caption{Accuracy (\%) Comparison of Wi-Fi and GNSS features- 30-second window.}
    \label{tab:wifi_results_30s}
    \begin{tabularx}{\columnwidth}{|p{2cm}|X|X|X|X|X|X|}
        \hline
        \textbf{Method} & \multicolumn{2}{c|}{\textbf{only GNSS}} & \multicolumn{2}{c|}{\textbf{only Wi-Fi}} & \multicolumn{2}{c|}{\textbf{GNSS + Wi-Fi}} \\
        \cline{2-7} 
        & \textbf{I} & \textbf{O} & \textbf{I} & \textbf{O} & \textbf{I} & \textbf{O} \\ \hline
        SVM & 69 & 100 & 51 & 87 & 70 & 100 \\ \hline
        RF & 79 & 99& 76 & 100 & 90 & 100 \\ \hline
        DT & 75 & 99 & 75 & 100 & 87 & 98 \\ \hline
    \end{tabularx}
\end{table}

\begin{table}[h!]
    \centering
    \caption{Accuracy (\%) Comparison of Wi-Fi and GNSS features- 60-second window.}
    \label{tab:wifi_results_60s}
    \begin{tabularx}{\columnwidth}{|p{2cm}|X|X|X|X|X|X|}
        \hline
        \textbf{Method} & \multicolumn{2}{c|}{\textbf{only GNSS}} & \multicolumn{2}{c|}{\textbf{only Wi-Fi}} & \multicolumn{2}{c|}{\textbf{GNSS + Wi-Fi}} \\
        \cline{2-7}
        & \textbf{I} & \textbf{O} & \textbf{I} & \textbf{O} & \textbf{I} & \textbf{O} \\ \hline
        SVM & 70 & 100 & 48 & 89 & 68 & 100 \\ \hline
        RF & 80 & 98 & 81 & 100 & 92 & 100 \\ \hline
        DT & 76 & 99 & 78 & 100 & 88 & 100 \\ \hline
    \end{tabularx}
\end{table}
From the comparison tables, we see that models trained on GNSS features only generally perform better than those trained on Wi-Fi data only. Combining both Wi-Fi and GNSS information leads to a notable improvement in accuracy, especially for the classification of indoor environments. 

\section{Practical Considerations}

\subsection{ML vs. Th-Based Methods}
We demonstrated that ML models generally outperform the Th-based model in terms 
of accuracy. The main reason for this difference is that the Th-based approach relies on a single feature (CNR per satellite) and treats each satellite independently, whereas ML models can exploit multiple features simultaneously and capture the underlying relationships between them within a timestamp. While this independence limits accuracy, the Th-based method also offers several advantages, particularly from a practical implementation perspective. 

The Th-based model relies on simple lookup tables to make decisions. 
This makes them highly scalable: for instance, adding a new satellite only 
requires updating the corresponding table entry, with little  
retraining. In contrast, ML models typically require retraining the entire system, 
which is substantially more time-consuming. Moreover, Th-based methods 
allow easy adjustment of the decision threshold to achieve a desired balance 
between error types, tailored to the needs of specific applications. Such 
adjustments can even be customized per satellite by modifying the lookup table. 
For ML models, by comparison, achieving similar goals generally requires 
designing new cost functions and retraining. Therefore, the choice between ML and Th-based approaches ultimately depends on the desired trade-off: ML models provide higher accuracy, while 
Th-based methods emphasize simplicity, scalability, and flexibility.

\begin{figure}[h!]
    \centering
    \includegraphics[width=0.65\columnwidth]{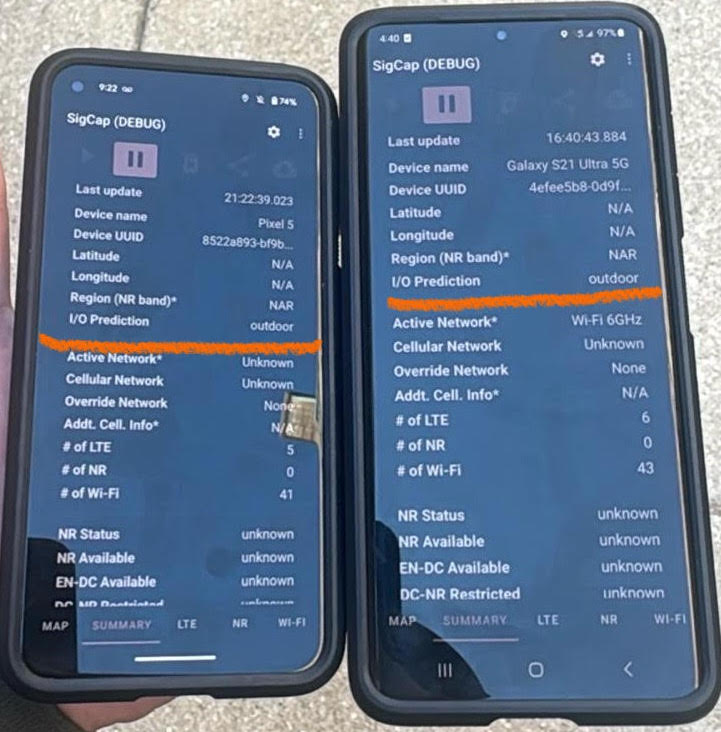} 
    \caption{SigCap application showing live I/O classification.}
    \label{fig:sigcap_app}
\end{figure}

A few selected models from this work are being integrated into the SigCap application to enable live I/O predictions. Figure~\ref{fig:sigcap_app} presents a screenshot of the demo version currently under development.

\subsection{Containment Instead of I/O }
\label{containment}
So far, the implicit assumption has been that accurately predicting the "human" environment (indoor vs. outdoor) and adjusting transmission power accordingly leads to optimal spectrum usage without introducing harmful interference. However, this assumption warrants scrutiny: Is the RF definition of indoor/outdoor the same as the human one? If it is not always valid, to what extent does it misrepresent interference risk?  

To proceed, it is necessary to clarify what constitutes an \textit{indoor} versus an \textit{outdoor} environment. Definitions vary across domains. For instance, \cite{Herneoja2023Outdoor} describes indoor spaces as those capable of maintaining a stable, conditioned climate, while the California Department of Public Health defines a structure as indoor if more than 50\% of its sides are closed by impermeable walls~\cite{CDPHOutdoorStructuresQA2021}. In general, an indoor environment is commonly perceived as a space enclosed by structural boundaries.

Nevertheless, these definitions can mislead in the context of Electromagnetic (EM) isolation. Two scenarios illustrate this issue:  
\begin{enumerate}
    \item A device classified as \textit{outdoor} may, in fact, be shielded from interference due to surrounding structures. For instance, a device located under an overpass bridge is technically outdoors but experiences significant EM isolation.  
    \item Conversely, a device located \textit{indoors} may be highly exposed to outdoor incumbents. An example is a device inside a modern airport terminal positioned near large glass windows, where emissions are weakly contained.  
\end{enumerate}

Such cases suggest that the conventional indoor--outdoor distinction is insufficient. As highlighted in \cite{FCCCharterLowPowerSensors2025}, a shift in focus is needed: from indoor/outdoor classification to containment. According to the FCC TAC report, containment measures ``the ability to confine Radio Frequency emission to specific boundaries, reducing harmful interference with spectrally or physically adjacent systems.'' In practice, a well-contained device is less likely to interfere with nearby systems (spatial containment) or with devices operating in adjacent bands (spectral containment).  

Although the FCC is still investigating a formal definition of containment, the concept offers a more direct and precise relationship between power control and interference mitigation than indoor/outdoor classification. Simply put, between two devices-regardless of whether they are indoors or outdoors-the one with stronger containment should be permitted higher transmit power.  Additionally, another key difference between I/O state and containment level is that containment can be measured (once defined) and reported as a non-binary value. In contrast to the I/O state, which is inherently binary, containment provides a spectrum of values that indicate the extent to which an environment is EM-contained. This enables more flexibility in defining transmit power limits. Instead of restricting devices to only two power regimes, flexible power levels can be assigned based on measured containment. For instance, within a building, containment levels may differ across interior areas, zones near windows, and outdoor spaces. Even different floors can exhibit varying levels of containment, as the top floor may be shielded by a single layer of walls while the ground floor is attenuated by multiple layers.  

\subsection{GNSS as an Illustration of Containment}

The FCC report suggests the use of sensors to monitor interference levels and spectrum usage, and to report this information to spectrum-sharing databases for dynamic allocation of spectrum resources. In this work, we use GNSS measurements as an illustrative case study to demonstrate how containment levels could be represented in practice.  

GNSS features-particularly carrier-to-noise ratio (CNR) and the number of observable satellites-are strongly correlated with the degree of EM isolation of a device relative to incoming satellite signals. Since satellites provide coverage from many directions at nearly all locations and times (albeit not perfectly uniformly), their signals can serve as a convenient probe of the surrounding environment. For example, if no satellite is detected from a particular direction, it likely indicates severe blockage along that path. This observation allows us to treat GNSS as a sensor that reflects the containment properties of different environments.  

To further explore this idea, we collected targeted measurements in 
environments that include both highly contained and poorly contained 
settings. 

\begin{figure}[ht]
    \centering
    \begin{subfigure}[b]{0.48\columnwidth}
        \centering
        \includegraphics[width=\linewidth]{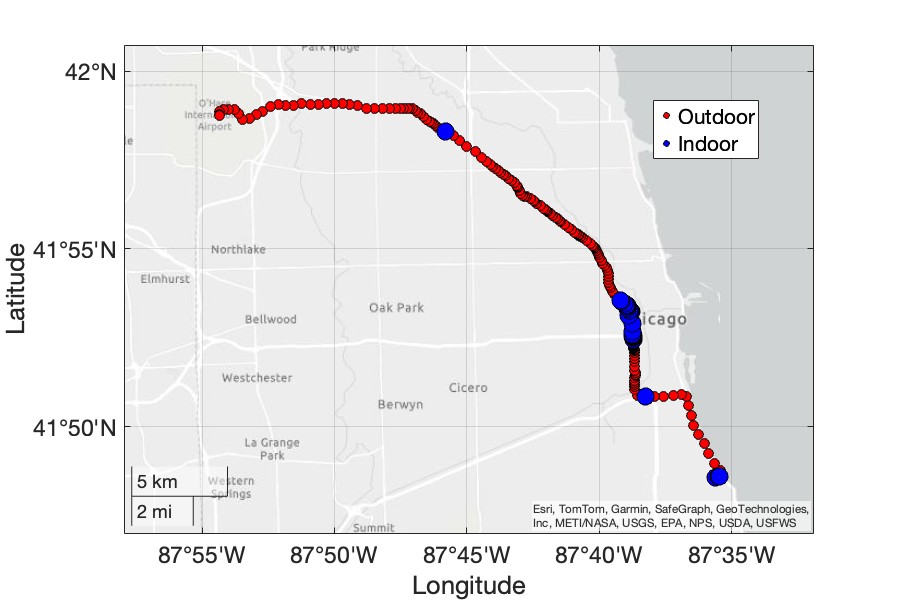}
        \caption{Driving}
        \label{fig:outdoor_contained_map}
    \end{subfigure}
    \hfill
    \begin{subfigure}[b]{0.46\columnwidth}
        \centering
        \includegraphics[width=\linewidth]{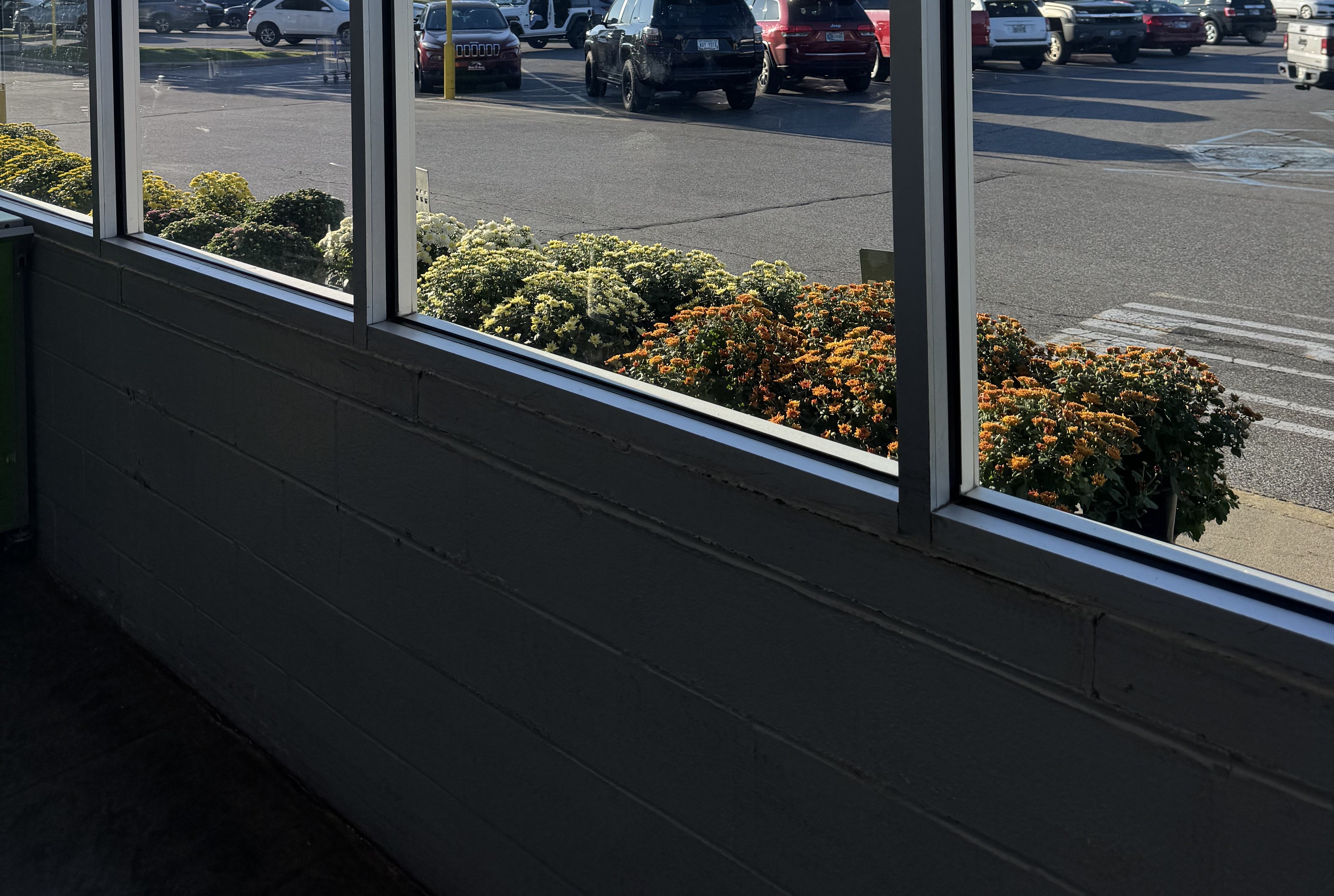}
        \caption{Grocery Store}
        \label{fig:indoor_contained_map}
    \end{subfigure}
    \caption{Geographic visualization of GNSS-based predictions for (a) an outdoor driving trip with under-bridge segments and (b) an indoor grocery store scenario.}
    \label{fig:io_contained_maps}
\end{figure}

\textbf{Measurement~1} (see Fig.~\ref{fig:outdoor_contained_map}) corresponds to a short outdoor 
driving trip from a residential building to O'Hare International Airport 
in Chicago, during which the vehicle passed under several bridges. These 
under-bridge segments represent poorly contained spaces. The average CNR 
and number of observed satellites, divided by containment level, are 
reported in Table~\ref{tab:containment_results}. The differences in values between well-contained 
and poorly contained segments are substantial. Interestingly, as shown 
in Fig.~\ref{fig:outdoor_contained_map} (blue marks), the vast majority of samples collected under bridges were 
classified as \emph{Indoor} by the model, despite being physically outdoors.

\textbf{Measurement~2} was conducted inside a grocery store in South Bend, 
with measurements taken both near large windows and in the interior aisles 
(see Fig.~\ref{fig:io_contained_maps}). Transition areas between these regions were not included, as their level of containment could not be reliably characterized. The 
corresponding averages are summarized in Table~\ref{tab:containment_results}. In this case, a substantial portion of the samples collected near windows were classified 
as \emph{Outdoor}, despite being indoors. 

\begin{table}[ht]
\centering
\caption{Comparison of well- and poorly-contained environments. The indoor represents inside a grocery store, while the outdoor represents the driving route to the airport. \% of correct prediction is according to the actual physical environment (indoor or outdoor).}
\label{tab:containment_results}
\resizebox{0.95\columnwidth}{!}{%
\begin{tabular}{l|c|c}
\toprule
\textbf{Containment Level} & \textbf{Outdoor (Driving)} & \textbf{Indoor (Grocery Store)} \\
\midrule
\multirow{5}{*}{Well-Contained}  
   & "Under the Bridges"      &  "Interior Sections"\\
    & Num. of samples: 130   &  Num. of samples: 53  \\
  & Avg. CNR (dB/Hz): 19.8 & Avg. CNR (dB/Hz): 22.8 \\
  & Avg. Num. of Observed Sat.: 22.8 & Avg. Num. of Observed Sat.: 19 \\
  & \% Predicted \textbf{Outdoor}: 10 & \% Predicted \textbf{Indoor}: 96 \\
\midrule
\multirow{5}{*}{Poorly-Contained}
  & "Not Under the Bridges"      &  "Near Large Windows"\\
   & Num. of Samples: 1100      &  Num. of Samples: 45\\
  & Avg. CNR (dB/Hz): 26.4 & Avg. CNR (dB/Hz): 24.8 \\
  & Avg. Num. of Observed Sat.: 32.8 & Avg. Num. of Observed Sat.: 29.7 \\
  & \% Predicted \textbf{Outdoor}: 100 & \% Predicted \textbf{Indoor}: 56 \\
\bottomrule
\end{tabular}}
\end{table}

Overall, these results highlight that GNSS directly reflects EM isolation, even at other frequencies, and that using I/O classification merely as a bridge to capture this relationship may be inefficient.


\section{Conclusion}

This work demonstrates that GNSS data effectively enhances the accuracy of I/O classification by utilizing ML and non-ML methods. Specifically, the RF model achieved high accuracy across a diverse dataset, while the threshold-based method, despite its simplicity, also exhibited reasonable performance. Our findings suggest that combining Wi-Fi features with GNSS data and employing temporal aggregation would yield very high accuracy, even in unseen locations with significantly different environments. It is also apparent that the human definition of "indoors" may not apply to EM signals, since locations near windows exhibit "outdoor" characteristics, which from an interference perspective is accurate. Our future work will seek to quantify “electromagnetic containment (‘EM indoor’)”. We also plan to extend this work by applying time series modeling and utilizing appropriate architectures, such as Recurrent Neural Networks (RNNs), to achieve improved accuracy.

\section*{Acknowledgements}
This research was funded in part by NSF Grants CNS-2229387, CNS-2346413, and AST-2132700.

\bibliographystyle{IEEEtran}
\bibliography{IEEEabrv,main}

\end{document}